\documentclass[12pt,pdftex]{article}
\usepackage{amsmath}
\usepackage{amssymb}
\usepackage[english]{babel}
\usepackage{abstract}
\usepackage{natbib}
\usepackage{amsthm}
\usepackage[title,titletoc,toc]{appendix}
\usepackage{graphicx} 
\usepackage{booktabs}
\usepackage{subcaption}
\usepackage{caption}
\usepackage{mathrsfs}
\usepackage{amsbsy}
\usepackage{enumerate}
\usepackage{comment}
\usepackage{enumitem}
\usepackage{sectsty}
\allowdisplaybreaks

\sectionfont{\fontsize{15}{15}\selectfont}
\subsectionfont{\fontsize{12}{15}\selectfont}

\newcommand{\bfu}{{\bf u}}

\newcommand{\bfx}{{\bf x}}

\newcommand{\bfA}{{\bf A}}
\newcommand{\bfB}{{\bf B}}

\newcommand{\bfI}{{\bf I}}

\newcommand{\bfX}{{\bf X}}
\newcommand{\bfY}{{\bf Y}}

\newcommand{\bfbeta}{\mbox{\boldmath $\beta$}}

\newcommand{\bftheta}{\mbox{\boldmath $\theta$}}

\newcommand{\bfmu}{\mbox{\boldmath $\mu$}}

\newsavebox\CBox

\newtheorem{theorem}{Theorem}
\newtheorem{corollary}{Corollary}

\newtheorem{lemma}{Lemma}
\newtheorem{proposition}{Proposition}

\begin{document}
\title{\large{\textbf{A Note on Nonlocal Prior Method in High Dimensional Setting}}}
\author{
	\textit{
			Yuanyuan Bian
			\footnote{University of Missouri, Columbia, MO, USA 
				.} 
		    , and
			Ho-Hsiang Wu
			\footnote{National Cancer Institute, Bethesda, MD, USA  
				.}
			}
		}
\date{\bigskip}

\maketitle

\begin{abstract}
We propose a new class of nonlocal prior to improve the performance of variable selection in high dimensional setting. We prove our new prior possesses the robustness to hyper parameter settings and is able to detect smaller decreasing signals.
\end{abstract}

\section{Introduction}\label{S:Int}
In this note we consider variable selection problem in high dimensional generalized linear models (GLMs). Let the data consists of a sequence of $\{\bfY_n, \bfX_n\}$, where $\bfY_n$, the responses, collect $n$ independent observations $y_i, i=1,...,n$, and $\bfX_n$, the regressors, forms $n\times p$ matrix. Given $\bfX_n$, each $y_i$ is conditionally independent and assumed to follow a distribution from the natural exponential family
\begin{equation}\label{model}
f(y_i|\theta_i)=c(y_i)\exp\left[y_i\theta_i-b(\theta_i)\right], i= 1,…,n,
\end{equation}
with natural parameter $\bftheta_n=\left[\theta_1,...,\theta_n\right]'$. The mean function $\bfmu=E(\bfY_n|\bfX_n)=b'(\bftheta_n)$. The matrix $\bfX_n$ influences $\bfY_n$ in the form of $\bftheta_n=\bfX_n\bfbeta$, where $\bfX_n$ can be further expressed as $[\bfx^{(1)},\bfx^{(2)},...,\bfx^{(p)}]$, a collection of regressors $\bfx^{(j)}$, for $j=1,...,p$, and $\bfbeta=[\beta_1,...,\beta_p]'$ denotes a $p\times1$ regression coefficient vector. This class of models include regression models whose responses are binomial, Poisson, and Gaussian with known variance.    

Now suppose the joint distribution of $\bfY_n$ is determined by a true parameter vector $\bfbeta_0\in\mathcal{R}^p$ supported on a small set $\mathcal{J}_0$, such that $\mathcal{J}_0\subset\left\{1,...,p\right\}$ and $\beta_{0j}\ne0$ if and only if $j\in\mathcal{J}_0$. Our interest here is to identify such set $\mathcal{J}_0$ that gives the most parsimonious true model. We are concerned with the asymptotic setting in which both $n$ and $p$ increase, particularly with $p$ being of substantial size with regard to $n$. Specifically, we assume that
\begin{enumerate}
	\item[(A1)] $\log(p)=O(n^{\omega})$ for some $\omega\in(0,\frac{1}{3}]$,
\end{enumerate}
which allows for subexponential growth of $p$ with respect to $n$. The variable selection problem under such assumption is challenging and recently draws great attention, see \cite{chen2012extended, liang2013bayesian} among others. Notably, many approaches resort to the sparsity assumption, by considering a priori bound, $q$, on the size of the models to be considered. By imposing such bound $q$, the total number of candidate models scales down dramatically from $2^p$ to $p^q$. Here we also assume the sparsity such that $|\mathcal{J}_0|\le q$ and 
\begin{itemize}
	\item[(A2)] $q\log(p)=O(n^{\psi})$ for some $\psi\in[0,\frac{1-\omega}{3})$,
\end{itemize}   
to allow $q$ to grow with $n$ but at a rather low rate.

We treat the variable selection problem with nonlocal prior method, a Bayesian variable selection framework that is first introduced by \cite{johnson2010use}. With the assignment of a nonlocal prior to induce probabilistic separation between the considered models, the nonlocal prior method has been proved to provide stronger parsimony than many of its competitors. \cite{johnson2012BS} modified the nonlocal prior method for Gaussian linear model, and proved its consistency in $p=O(n)$ setting. Recently, \cite{shin2015scalable} undertook the study in $p=O(e^n)$ setting, and not only established the consistency results (with other conditions), but also demonstrated several advantages of their nonlocal prior method against other current-state-of-art methods, such as $g$-prior method and penalized likelihood approaches.

Despite the recent advances of nonlocal prior method, one aspect, i.e., the specification of hyper parameters of the prior, remains an open research interest. To this end, we go back to Shin's nonlocal prior method with a new class of nonlocal prior, and set out to demonstrate its potential advantages including robustness to hyper parameter settings and the ability in detecting small decreasing signal.

\section{Main Results}\label{S:Main}
\subsection{Set up and notation}
Recall the problem of interest here is to recover the support of $\bfbeta_{0}$, that is, the set 
\begin{equation*}
\mathcal{J}_0:\left\{j\in[p]: \beta_{0j}\ne 0\right\},
\end{equation*} where $|\mathcal{J}_0|\le q$ and $[p]:=\left\{1,...,p\right\}$. We first introduce some notations for submodel. Let $\mathcal{J}$, such that $\mathcal{J}\subset[p]$, index a generic submodel consisting of a subset of $p$ covariates. Let $\bfbeta_{\mathcal{J}}$ denote the corresponding length $|\mathcal{J}|$ regression coefficient vector of this submodel. Let $\hat\bfbeta_{\mathcal{J}}$ denote the MLE of $\bfbeta_{\mathcal{J}}$, and $\hat\bfbeta_{\mathcal{J},\scriptscriptstyle{\text{PM}}}$ denote the posterior mode of~$\bfbeta_{\mathcal{J}}$. We simply write $\mathbb{R}^{|\mathcal{J}|}$ for the parameter space comprising all $\bfbeta$ with length $|\mathcal{J}|$. We further denote the log-likelihood function by $\ell(\cdot)$, the score function by $\mathcal{S}(\cdot)$, the negative Hessian of the log-likelihood function by $\mathcal{H}(\cdot)$, and the negative Hessian of the log-posterior density function by $\mathcal{H}^*(\cdot)$.

We now describe the nonlocal prior method in \cite{shin2015scalable}. Given a generic model $\mathcal{J}$, they first assigned on the regression coefficients $\bfbeta_{\mathcal{J}}$ the product inverse moment prior (piMOM, \citealt{johnson2012BS}), 
$$
\pi(\bfbeta_{\mathcal{J}}|r, \tau)=\frac{\tau^{r|\mathcal{J}|/2}}{\Gamma(r/2)^{|\mathcal{J}|}}\prod_{j=1}^{|\mathcal{J}|}|\beta_{\mathcal{J}j}|^{-(r+1)}\exp(-\tau/\beta^2_{\mathcal{J}j}), 
$$
and then carried out the marginal likelihood $\mathcal{M}_{\mathcal{J}}$, using the standard Laplace approximation.
To incorporate the prior believe on sparsity, they assumed on model $\mathcal{J}$ a uniform prior 
$
\pi_{\mathcal{J}}\propto\bfI(|\mathcal{J}|\le q),
$ so that the model space is restricted to models whose size is less than or equal to the upper bound~$q$. Note $\bfI(\cdot)$ above denotes the indicator function. Putting all these together, \cite{shin2015scalable} then computed the model posterior probability for every model in consideration, and then used this quantity as criterion to identify $\mathcal{J}_0$ with the highest posterior model.  

Following \cite{shin2015scalable}, we find that the model posterior probability of the true model $\mathcal{J}_0$ can be expressed in the form of
\begin{equation}\label{modelpost}
p(\mathcal{J}_0|\bfY_n)=\left[1+\sum_{\mathcal{A}}{\frac{\mathcal{M}_{\mathcal{J}}}{\mathcal{M}_{\mathcal{J}_0}}}+\sum_{\mathcal{B}}{\frac{\mathcal{M}_{\mathcal{J}}}{\mathcal{M}_{\mathcal{J}_0}}}\right]^{-1},
\end{equation}
where $\mathcal{A}$ denotes the set $\left\{\mathcal{J}: \mathcal{J}_0\subset\mathcal{J}, |\mathcal{J}|\le q\right\}$ that collects nested models, and $\mathcal{B}$ denotes the set $\left\{\mathcal{J}: \mathcal{J}_0\not\subseteq\mathcal{J}, |\mathcal{J}|\le q\right\}$ that collects non-nested models. Here it is clear to see that the last two summations play crucial roles, since if they go to zero as the sample size $n$ increases, then the method achieves the Bayesian variable selection consistency \citep{bayarri2012criteria}. In the literature, it is recognized that the most appealing contribution of nonlocal prior method is the improvement, when compared with other Bayesian methods that employ local priors, of the variable selection within the set $\mathcal{A}$. The convergence rate (toward zero) of the ratio of $\mathcal{M}_{\mathcal{J}}$ versus $\mathcal{M}_{\mathcal{J}_0}$ within $\mathcal{A}$ is sub-exponential with Shin's nonlocal prior method, but is polynomial with local prior method.      

To see this, we first introduce below two assumptions on the Hessian $\mathcal{H}(\cdot)$:
\begin{itemize}
	\item[(B1)] There exist constants $c_{L}$ and $c_{U}$ such that for all $|\mathcal{J}|\le q$ and all $\bfbeta\in\mathbb{R}^{|\mathcal{J}|}$, the negative Hessian function, $\mathcal{H}(\bfbeta)$, is properly bounded as
	$
	c_{L}\bfI_{|\mathcal{J}|} \preceq n^{-1}\mathcal{H}(\bfbeta) \preceq c_{U}\bfI_{|\mathcal{J}|}, 
	$
	where the notation ``$\preceq$'' refers to the ordering with $\bfA\preceq\bfB$ whenever $\bfA-\bfB$ is positive semidefinite.
	\item[(B2)] There is a constant $c_{D}$ such that 
	$
	n^{-1}||\mathcal{H}(\bfbeta)-\mathcal{H}(\bfbeta^*)||_S\le c_{D}\cdot||\bfbeta-\bfbeta^*||_2
	$
	for all $|\mathcal{J}|\le q$ and all $\bfbeta, \bfbeta^*\in\mathbb{R}^{|\mathcal{J}|}$, where $||\cdot||_S$ is the spectral norm of a matrix.
\end{itemize}
These two assumptions are concerned with the asymptotic identifiability. In other words, they ensure that with large sample size, the true model is always properly bounded away from the wrong model. Next, we consider a proposition to see the asymptotic behavior of $\sum_{\mathcal{A}}\mathcal{M}_\mathcal{J}/\mathcal{M}_{\mathcal{J}_0}$ in (\ref{modelpost})
\begin{proposition}\label{LemmalogM}
	Suppose \textnormal{piMOM} is assumed. Fix $r, \epsilon,$ and $\nu>0$. If regularity assumptions \textnormal{(A1), (A2), (B1)} and \textnormal{(B2)} are satisfied, then for all $\mathcal{J}$ such that $\mathcal{J}_0\subset\mathcal{J}, |\mathcal{J}|\le q$, we have
	\begin{equation}\label{logM}
		\log\left(\frac{\mathcal{M}_\mathcal{J}}{\mathcal{M}_{\mathcal{J}_0}}\right)\asymp (1+\epsilon)\log p^{(\nu+|\mathcal{J}|-|\mathcal{J}_0|)}-
		\left(
		\sum_{j\in\mathcal{J}}\tau\hat\beta^{-2}_{\mathcal{J} i,\scriptscriptstyle{\text{PM}}} -
		\sum_{j\in\mathcal{J}_0}\tau\hat\beta^{-2}_{\mathcal{J}_0 i,\scriptscriptstyle{\text{PM}}}
		\right)
	\end{equation}
	where the notation $``\asymp$'' refers to the asymptotic equivalence.
\end{proposition}
The proof is straightforward and thus omitted. In the right hand side of (\ref{logM}), the first term comes from the log-likelihood ratio, while the second term (the two summations within the bracket) comes from the log-prior ratio. The convergence rate of the log-likelihood ratio is established in the Theorem 2.2 of \cite{barber2015high} and holds here since they assume more general regularity conditions (at cost of more tedious technical work). From Proposition 1, we see two substantial factors determine the convergence rate of the log-prior ratio, one is the exponential kernel of piMOM, and another is the convergence rate (toward zero) of the posterior mode $\hat\bfbeta_{\mathcal{J}_0,\scriptscriptstyle{\text{PM}}}$. With additional regularity conditions and properly chosen $\tau$, we can show that the convergence rate of the log-prior ratio is of order $\mathcal{O}(n^c)$ for some $c>0$ such that the summation $\sum_{\mathcal{A}}\mathcal{M}_\mathcal{J}/\mathcal{M}_{\mathcal{J}_0}$ in (\ref{modelpost}) converges toward zero exponentially fast. 

Note although Proposition \ref{LemmalogM} elucidates the advantage of using nonlocal prior method in an asymptotic sense, it does not provide practical guide on setting hyper parameter $\tau$ since the proportionality is unknown.    

\subsection{A new class of nonlocal prior and asymptotic results}
For practical purpose, \cite{nikooienejad2016bayesian} have proposed for setting $\tau$ a heuristic procedure that is lately shown by \cite{shin2015scalable} to work well with piMOM in the high-dimensional setting. Below we propose a new class of robust nonlocal prior and indicate why we think it may be of use with Nikooienejad's procedure.
 
\begin{proposition} Suppose we replace $\tau$ in \textnormal{piMOM} with $\tau_j$ and further assume each $\tau_j$ follows an inverse-gamma distribution with shape $(r+1)/2$ and scale $\lambda$, then we have
\begin{equation}
\begin{split}
\pi(\bfbeta_{\mathcal{J}}|r, \lambda)
&=\int...\int\pi(\bfbeta_{\mathcal{J}}|r, \tau_j)\pi(\tau_j|\lambda) d\tau_1...d\tau_{|\scriptscriptstyle{\mathcal{J}}|}\\
&=\prod_{j=1}^{|\mathcal{J}|}\frac{|\beta_{\mathcal{J}j}|^{-(r+1)}\lambda^{(r+1)/2}\sqrt{\pi}}{2\Gamma(\frac{r+1}{2})\Gamma(\frac{r}{2})\sqrt{\lambda}}\exp\left(-2\sqrt{\frac{\lambda}{\beta^2_{\mathcal{J}j}}}\right).
\end{split}
\end{equation}	
	
\end{proposition} 

\indent\textbf{Remark 1.} We call this prior scale mixture piMOM (spiMOM), as we introduce a hyper prior on piMOM's scale parameter. By considering an additional layer in the hierarchy, we take into account the uncertainty of $\tau$. Consequently, spiMOM has a sub-exponential kernel, which leads to a heavier tail and more flat spikes around the origin, when compared with piMOM. Note $\lambda$ has less impact than $\tau$ on determining the minimum value of $\bfbeta$ to be considered as non-trivial. To determine the value of $\lambda$, the Nikooienejad's procedure can be applied. Essentially, as the Nikooienejad's procedure involves random sampling among models, the robustness of the spiMOM is appealing. 

Another advantage of spiMOM, as will be seen later, is its ability to detect smaller decreasing signal. Below we establish the asymptotic behavior of the posterior mode under spiMOM. We first begin with additional conditions and a lemma on the MLE.

\begin{itemize}
	\item[(C1)] For all $i \in \left\{1,...,n\right\}$ and $j\in\mathcal{J}\supseteq\mathcal{J}_0$, $x_{ij}\left[y_i-b'(\bfx_i^T\bfbeta_0)\right]=\mathcal{O}(1)$.
	
    \item[(C2)] $\min_{j\in\mathcal{J}_0}\{\beta_{0j}\}=\mathcal{O}(n^{-m})$, $m>0.$
\end{itemize}

\begin{lemma}\label{LemmaMLE} Suppose for all $\mathcal{J}\supseteq\mathcal{J}_0$ with $|\mathcal{J}|\le q$ the conditions \textnormal{(A1), (A2), (B1), (B2), (C1)} and \textnormal{(C2)} hold. Then 
$
	||\hat\bfbeta_{\mathcal{J}}-\bfbeta_{0}||_2=\mathcal{O}(n^{-1/3}).
$		
\end{lemma}
	
\indent\textbf{Remark 2.} Here Lemma \ref{LemmaMLE} is crucial in that it reveals the asymptotic lower bound of the posterior mode. To see this, note that the nonlocal prior is symmetric around the origin , hence resulting posterior density has posterior model (global mode) occuring at the same orthant in $\mathbb{R}^{|\mathcal{J}|}$ as the MLE, and has many other local modes in other orthants. Essentially, because of the convexity of the posterior density by which the posterior mode is contracted toward the MLE, we have the posterior mode converge toward the truth $\bfbeta_0$
	
\begin{theorem}\label{Main}
Suppose all the conditions of Lemma \ref{LemmaMLE} hold and \textnormal{spiMOM} is assigned. Then, for any $\epsilon^*_n\succ(r\lambda/n)^{1/3}$, the posterior mode $\hat\bfbeta_{\mathcal{J},\scriptscriptstyle{\text{PM}}}$ satisfies 
$$
p\left[\hat\bfbeta_{\mathcal{J},\scriptscriptstyle{\text{PM}}}\notin\mathcal{N}(\hat\bfbeta_{\mathcal{J}};\epsilon^*_n)\right]\rightarrow 0,
$$
where $\mathcal{N}(\textbf{u};\epsilon)=\left\{\textbf{v}\in\mathbb{R}^{|\mathcal{J}|}:|v_j-u_j|\le\epsilon, j\in\mathcal{J}\right\}$.
\end{theorem}	

\indent\textbf{Remark 3.} Theorem \ref{Main} shows that under regularity conditions, the maximum a posterior estimator derived from spiMOM is asymptotically within $(r\lambda/n)^{1/3}$-neighborhood of the MLE. While \cite{shin2015scalable} have proved that the maximum a posterior estimator obtained from piMOM resides at a distance of $(\tau/n)^{1/4}$ from the MLE (assuming fixed $r$), here the implication of Theorem \ref{Main} is apparent: the use of spiMOM improves the nonlocal prior method in detecting small decreasing coefficients.

\indent\textbf{Remark 4.} It can be found in the proof of Theorem \ref{Main} that increasing $r$ (with sample size $n$) though increases the penalty on complex models, but impedes the nonlocal prior method from detecting small coefficients. Therefore, to strike a balance we suggest fix $r$ as constant.

Finally, we state below that spiMOM achieves Bayesian variable selection consistency under the proposed regularity conditions.  	
\begin{corollary}\label{Main2}
	 Suppose for all $\mathcal{J}$ with $|\mathcal{J}|\le q$ the conditions \textnormal{(A1), (A2), (B1), (B2), (C1)} and \textnormal{(C2)} hold and \textnormal{spiMOM} is assigned. Fix $r, \epsilon, \nu>0$, and $0<m<\frac{1}{3}$. Then there exist constants $c_1$ and $c_2$ such that if $c_1(1+\epsilon)(1+\nu)<\lambda^{1/6}<<c_2n^{2/9}$, then  $p(\mathcal{J}_0|\bfY_n)\xrightarrow{p}1.$
\end{corollary}

\section{Proofs}\label{S:pfs}
\subsection{Proof of Proposition 2}
Without loss of generality, we present the proof for univariate $\beta$. The extension to multivariate $\bfbeta$ is straightforward due to the assumption of independence. Now with $\pi(\beta|r, \tau)$ and $\tau$ following inverse-gamma with shape $(r+1)/2$ and scale $\lambda$ we can show    
\begin{align*}
\pi(\beta|r ,\lambda)
&=\int_{\mathbb{R}^+}\pi(\beta|r, \tau)\pi(\tau|  (r+1)/2 , \lambda)\text{d}\tau\\
&=\int_0^\infty  
\frac{\tau^{\frac{r}{2}}}{\Gamma(\frac{r}{2})}|\beta|^{-(r+1)}\exp\left(-\frac{\tau}{\beta^2}\right)
\frac{\lambda^\frac{r+1}{2}}{\Gamma(\frac{r+1}{2})}\tau^{-\frac{r+1}{2}-1}\exp(-\frac{\lambda}{\tau})\text{d}\tau\\
&=\frac{|\beta|^{-(r+1)}\lambda^\frac{r+1}{2}}{\Gamma(\frac{r}{2})\Gamma(\frac{r+1}{2})}\exp\left(-2\sqrt{\frac{\lambda}{\beta^2}}\right)\int_0^\infty 2\exp\left(-\lambda t^2-(\sqrt{\beta^2}t)^{-2}\right)\text{d}t \stepcounter{equation}\tag{\theequation}\label{myeq1}\\
&=\frac{|\beta|^{-(r+1)}\lambda^\frac{r+1}{2}}{\Gamma(\frac{r}{2})\Gamma(\frac{r+1}{2})}\exp\left(-2\sqrt{\frac{\lambda}{\beta^2}}\right)\int_0^\infty \frac{1}{\sqrt{\lambda}}\exp\left(-\iota^2\right)\text{d}\iota \stepcounter{equation}\tag{\theequation}\label{myeq2}\\
&=\frac{|\beta|^{-(r+1)}\lambda^\frac{r+1}{2}\sqrt{\pi}}{2\Gamma(\frac{r}{2})\Gamma(\frac{r+1}{2})\sqrt{\lambda}}\exp\left(-2\sqrt{\frac{\lambda}{\beta^2}}\right).
\end{align*}
Note that the Equation (\ref{myeq1}) is obtained using change of variable with $\tau=t^{-2}$. The Equation (\ref{myeq2}) results from Cauchy-Schl$\text{\"o}$milch transformation
$$
\int_0^\infty g\left\{\left(at-bt^{-1}\right)^2\right\}\text{d}t=\frac{1}{2a}\int_0^\infty g\left(\iota^2\right)\text{d}\iota. 
$$   

\subsection{Proof of Lemma 1}
Since condition (C2) is satisfied, for any unit vector $\bfu\in\mathbb{R}^{|\mathcal{J}|}$,  we can set $\bfbeta_{0}=\bfbeta+n^{-m}\bfu$, $m>0$. It is straightforward to see that for sufficiently large $n$, $\bfbeta$ falls into the neighborhood of $\bfbeta_0$ so that condition (B1) and (B2) apply. Therefore we have
$$
\ell(\bfbeta_{\mathcal{J}})-\ell(\bfbeta_0)\le n^{-m}\bfu^T\mathcal{S}_{\mathcal{J}}(\bfbeta_0)-c(1-\epsilon)n^{1-2m}
$$
for all $\mathcal{J}\supseteq\mathcal{J}_0$ with $|\mathcal{J}|\le q$, which implies that
$
P\left\{\ell(\bfbeta_{\mathcal{J}})-\ell(\bfbeta_0)>0\right\}\le\sum_{j\in\mathcal{J}}P(\mathcal{S}^2_{\mathcal{J} j}(\bfbeta_0)\ge n^{2-2m})
$, where $\mathcal{S}_{\mathcal{J} j}$ denotes the $j^{th}$ element of $\mathcal{S}_{\mathcal{J}}$. Note that to show the MLE is consistent with the desired convergence rate, it suffices to show that $\sum_{j\in\mathcal{J}}P(\mathcal{S}_{\mathcal{J} j}(\bfbeta_0)\ge n^{1-m})$ converges toward zero.

Now since the condition (C1) is satisfied, we can make use of Benette's inequality and have
$ 
P(\mathcal{S}_{\mathcal{J} j}(\bfbeta_0)\ge n^{1-m})\le\exp\left[-n^{1-2m}/(2+\mathrm{o}(1))\right].
$ 	
Observe that $|\mathcal{J}|$ is no more than $p^{q}\le\exp\left\{\mathcal{O}(n^{1/3})\right\}$. Therefore, we have the MLE exist and fall within the $n^{-1/3}$-neighborhood of $\bfbeta_0$. The lemma is proved.

\subsection{Proof of Theorem 1}
First note that the log-prior density of the nonlocal prior we considered here takes the following general form
$
\log\pi(\bfbeta_{\mathcal{J}})\propto -r\sum_{j=1}^{|\mathcal{J}|}\log(\beta_{\mathcal{J} j}^2)-\sum_{j=1}^{|\mathcal{J}|}\left\{\varphi\beta_{\mathcal{J} j}^{-2}\right\}^{\zeta},
$
such that $(\zeta, \varphi)=(1, \tau)$ corresponds to the piMOM, and $(\zeta, \varphi)=(1/2, \lambda)$ corresponds to the spiMOM, respectively. Next note that, by applying second order Taylor expansion on the log-likelihood density around the MLE, we arrive at
$
\ell(\bfbeta_{\mathcal{J}})=\ell(\hat\bfbeta_{\mathcal{J}})-\frac{1}{2}\gamma^T\mathcal{H}_{\mathcal{J}}(\bfbeta^*_{\mathcal{J}})\gamma,
$  
where $\bfbeta^*_{\mathcal{J}}=\hat\bfbeta_{\mathcal{J}}+\xi\gamma$, $\gamma=\bfbeta_{\mathcal{J}}-\hat\bfbeta_{\mathcal{J}}$, and $\xi\in\left[0,1\right]$. Putting together the log-prior density and log-likelihood, we derive the score function, i.e., the first order partial derivative of unnormalized log-posterior density with respect to $\bfbeta_{\mathcal{J}}$ as
$
\mathcal{S}^*(\bfbeta_{\mathcal{J}})=-\mathcal{H}_{\mathcal{J}}(\bfbeta^*_{\mathcal{J}})\gamma-r\bfbeta_{\mathcal{J}}^{\wedge(-1)}+\varphi\zeta\bfbeta_{\mathcal{J}}^{\wedge(-1-2\zeta)},
$
where we write $\bfbeta_{\mathcal{J}}^{\wedge(-1)}$ to raise each element of $\bfbeta_{\mathcal{J}}$ to power of minus one. 

From the score function we find that, conditioning on (B1) and (B2) being true, each element of mode must satisfy
$
-an(\beta_{j}-\hat\beta_{j})-r\beta_{j}^{-1}+\varphi\zeta\beta_{j}^{-1-2\zeta}=0  
$ for some constant $a>0$. Note here we drop the subscription of model index for simple exposition. Eventually, as every element of mode is nontrivial, we have 
\begin{equation}\label{elem_score}
 \frac{n}{r\varphi}(\beta_{j}-\hat\beta_j)\beta_{j}^{1+2\zeta}\xrightarrow{p}c
\end{equation}
for some constant $c$. 

Without loss of generality, we consider the two modes (global and local modes) of a generic element $\beta_{j}$ occurring given that sign$(\hat\beta_{j', \scriptscriptstyle{\text{PM}}})=$sign$(\hat\beta_{j'})$ for all $j'\ne j, j'\in\mathcal{J}$. For convenient notations, we write $\beta_{0j}$ for true value that corresponds to such element $\beta_{j}$, and write $\check{\beta}_1$ for global mode (i.e. posterior mode) of $\beta_{j}$ whose sign is the same with its MLE, and $\check{\beta}_2$ for local mode whose sign is different from the MLE. We simply write $c$ for generic constant if there is no confusion.  

Consider first the case of fixed $\beta_{0j}\ne0$. First note $\hat\beta_{j}\xrightarrow{p}\beta_{0j}$ implies that the global mode also converges to a constant, and thus $\check{\beta}_1^{1+2\zeta}\xrightarrow{p}c$ and $n(\check{\beta}_1-\hat\beta_{j})/(r\varphi)\xrightarrow{p}c$. For the local mode, since $(\check{\beta}_2-\hat\beta_{j})\xrightarrow{p}c$, we have $n\check{\beta}^{1+2\zeta}_2/(r\varphi)\xrightarrow{p}c$. 

Now consider the case of $\beta_{0j}=0$. First note that both modes converge toward zero at the convergence rate no faster than $n^{-1/3}$. To see this, consider $\check{\beta}_1=\mathcal{O}(n^{-1/3-\epsilon^*})$ with some $\epsilon^*>0$, then it follows that $ n(\check\beta_1-\hat\beta_j)\check\beta_1^{1+2\zeta}/(r\varphi)\asymp n^{2/3}\check{\beta}_1^{1+2\zeta}/(r\varphi)$ does not converge to some constant as required, a contradiction regardless of $\zeta=1$ or $1/2$. In contrast, consider $\check{\beta}_1=\mathcal{O}(\left[r\varphi/n\right]^{1/3-\epsilon^*})$, then it follows that $n\check{\beta}_1^{2+2\zeta}/(r\varphi)\xrightarrow{p}c$ and therefore $\check{\beta}_1=\mathcal{O}(\left[r\varphi/n\right]^{1/4})$ if $\zeta=1$ or $\check{\beta}_1=\mathcal{O}(\left[r\varphi/n\right]^{1/3})$ if $\zeta=1/2$. Finally, observing that $(\check{\beta}_2-\hat{\beta}_j)\asymp\check{\beta}_2$ (due to the opposite signs), we reach the same conclusion of the convergence rate for $\check{\beta}_2$. 

Finally consider the case when $\beta_{0j}= O(n^{-m})$. We first focus on the global mode.  If $\mathcal{O}(|\hat\beta_{j}-\beta_{0j}|)\preceq\mathcal{O}(n^{-m})\prec\mathcal{O}(\left[r\varphi/n\right]^{1/(2+2\zeta)})$, a similar argument to the case $\beta_{0j}=0$ implies that $\check{\beta}_1$ converges toward $\beta_{0j}$ but strictly slower than $O(n^{-m})$ such that $n\check{\beta}^{2+2\zeta}_1(1-\hat\beta_{j}/\check{\beta}_1)/(r\varphi)$ converges to a finite constant as required. However, if $\mathcal{O}(n^{-m})\succ\mathcal{O}(\left[r\varphi/n\right]^{1/(2+2\zeta)})$ we shall see that assuming  $|\check\beta_1-\hat\beta_{j}|\succ\mathcal{O}(\left[r\varphi/n\right]^{1/(2+2\zeta)})$ leads to a contradiction. Note that if $ |\check\beta_1-\hat\beta_{j}|\succ\mathcal{O}(\left[r\varphi/n\right]^{1/(2+2\zeta)})$, then $|\check\beta_1-\hat\beta_{j}|=\mathcal{O}(\check\beta_1)$ and thus $n(\check\beta_1-\hat\beta_j)\check\beta_1^{1+2\zeta}/(r\varphi)=\mathcal{O}(n\check{\beta}_1^{2+2\zeta}/(r\varphi))$, which gives that $\mathcal{O}(\check\beta_1)\succ\mathcal{O}(\left[r\varphi/n\right]^{1/(2+2\zeta)})$ could not yield required results by (\ref{elem_score}), leading to a contradiction. Now consider the local mode. Similarly, if 
$\mathcal{O}(n^{-m})\prec\mathcal{O}(\left[r\varphi/n\right]^{1/(2+2\zeta)})$, we have $\check{\beta}_2-\hat\beta_{j}=\mathcal{O}(\left[r\varphi/n\right]^{1/(2+2\zeta)})$ as the signs differ, thus we have $n\left[\varphi/n\right]^{1/(2+2\zeta)}\check{\beta}^{1+2\zeta}_2\xrightarrow{p}~c$. If $\mathcal{O}(n^{-m})\succ\mathcal{O}(\left[r\varphi/n\right]^{1/(2+2\zeta)})$, we have $\check{\beta}_2-\hat\beta_{j}=\mathcal{O}(n^{-m})$ and hence we have $n^{1-m}\check{\beta}_2^{1+2\zeta}\xrightarrow{p}c$.  

Overall, we have shown that in any case, there is a high probability for posterior mode to fall within the $(\varphi/n)^{1/({2+2\zeta})}$-neighborhood of the MLE. Besides, the posterior mode of spiMOM holds faster convergence rate than that of piMOM. On the other hand, the local mode also converges toward the true value, but at a rather lower rate. Finally, we observe that increasing $r$ or $\varphi$ (with respect to $n$) impedes the convergence rate toward the true value. However, as $\varphi$ is in the exponential kernel of piMOM (or spiMOM) while $r$ is not, we see from Proposition 1 that a fixed $r$ is preferred.        

\subsection{Proof of Corollary 1}
To prove the consistency,  we will show that the last two summations of (\ref{modelpost}) decrease to zero as sample size increases. Below we discuss these two cases.

\vspace{3mm}
\noindent\textbf{Case I: $\mathcal{J}\in\mathcal{A}$}\\ 
Now that spiMOM is assigned, we can rewrite the Equation (\ref{logM}) as
\begin{equation}\label{GenLogM}
	\log\left(\frac{\mathcal{M}_\mathcal{J}}{\mathcal{M}_{\mathcal{J}_0}}\right)\asymp (1+\epsilon)\log p^{(\nu+|\mathcal{J}\setminus\mathcal{J}_0|)}-
\left(
\sum_{j\in\mathcal{J}}\left\{\lambda\hat\beta^{-2}_{\mathcal{J}j,\scriptscriptstyle{\text{PM}}}\right\}^{1/2} -
\sum_{j\in\mathcal{J}_0}\left\{\lambda\hat\beta^{-2}_{\mathcal{J}_0j,\scriptscriptstyle{\text{PM}}}\right\}^{1/2}
\right).
\end{equation}  
From the condition (A1), we have the first term in (\ref{GenLogM}) such that
\begin{equation*}
(1+\epsilon)\log p^{(\nu+|\mathcal{J}\setminus\mathcal{J}_0|)}\preceq(|\mathcal{J}\setminus\mathcal{J}_0|)(1+\epsilon)(1+\nu)n^{1/3}.
\end{equation*} 
From the result of Theorem \ref{Main}, we have the last term in (\ref{GenLogM}) dominated by $\sum_{j\in\mathcal{J}\setminus\mathcal{J}_0}\left\{\varphi\hat\beta^{-2}_{ \mathcal{J}j,\scriptscriptstyle{\text{PM}}}\right\}^{1/2}$ such that 
\begin{equation*}
\sum_{j\in\mathcal{J}\setminus\mathcal{J}_0}\left\{\varphi\hat\beta^{-2}_{ \mathcal{J}j,\scriptscriptstyle{\text{PM}}}\right\}^{1/2}\asymp(|\mathcal{J}\setminus\mathcal{J}_0|)\mathcal{O}(\lambda^{1/6} n^{1/3}).
\end{equation*} 
Finally, we see that if $\epsilon$ and $\nu$ are sufficiently small such that $\lambda^{1/6}>c(1+\epsilon)(1+\nu)$, then we have $\sum_{\mathcal{A}}\mathcal{M}_\mathcal{J}/\mathcal{M}_{\mathcal{J}_0}$ converges toward zero.      

\vspace{3mm}
\noindent\textbf{Case II: $\mathcal{J}\in\mathcal{B}$}\\
We first cite the result of Theorem 2.2 of \cite{barber2015high} which states that 
\begin{equation}
	\ell(\hat\bfbeta_{\mathcal{J}_0})-\ell(\hat\bfbeta_{\mathcal{J}})\succeq c n \min_{j\in\mathcal{J}_0}|\beta_{0j}|^2\asymp\mathcal{O}(n^{1-2m}). 
\end{equation}
Note that their result holds here because they assume more general regularity conditions. Now observe that the log-prior ratio is bounded from below by 
$-\sum_{j\in\mathcal{J}_0}\left\{\lambda\hat\beta^{-2}_{\mathcal{J}_0j,\scriptscriptstyle{\text{PM}}}\right\}^{1/2}$. Given $m<1/3$, we have $\min_{j\in\mathcal{J}_0}|\hat\beta_{0j,\scriptscriptstyle{\text{PM}}}|^2=\mathcal{O}(n^{-m})$, and thus we have  
$$-\sum_{j\in\mathcal{J}_0}\left\{\lambda\hat\beta^{-2}_{\mathcal{J}_0j,\scriptscriptstyle{\text{PM}}}\right\}^{1/2}\succeq-c\lambda^{1/2}n^{-m}.$$
As a consequence, as long as $\lambda^{1/2}\prec n^{2/3}$, we have $\log(\mathcal{M}_{\mathcal{J}_0}/\mathcal{M}_{\mathcal{J}})$ dominated by $\mathcal{O}(n^{1-2m})\succ O(n^{1/3})$ and thus arrive at the conclusion that  $\sum_{\mathcal{B}}\mathcal{M}_\mathcal{J}/\mathcal{M}_{\mathcal{J}_0}$ converges toward zero.

\section{Conclusion}\label{S:sim}
In this note we discuss spiMOM, a new class of nonlocal prior that holds the robustness to specification of hyperparameters. Under certain regularity conditions, spiMOM provides maximum a posterior estimate converging at the same optimal rate as the MLE toward the truth. Overall, our approach may prove especially useful in applications of detecting small decreasing signal in high dimensional sparse data.      

\setcounter{secnumdepth}{0}
 
\baselineskip=12pt \vskip 2mm\noindent
\bibliographystyle{jasa}
\bibliography{Bib_Reference}
\end{document}